\documentclass[superscriptaddress, amsmath,amssymb, aps, prb, twocolumn]{revtex4-2}
\usepackage{graphicx}
\usepackage{dcolumn}
\usepackage{bm}
\usepackage{xcolor}
\usepackage{comment}
\newcommand{\knp}{KNiPO$_4$}
\begin{document} 
\title{Revisiting the magnetic and crystal structure of multiferroic KNiPO$_4$}
\author{Alexandre Pages}
\affiliation{Institute  of  Physics,  \'Ecole  Polytechnique  F\'ed\'erale  de  Lausanne  (EPFL),  CH-1015  Lausanne,  Switzerland}%
\author{Jian-Rui Soh}%
\affiliation{Institute  of  Physics,  \'Ecole  Polytechnique  F\'ed\'erale  de  Lausanne  (EPFL),  CH-1015  Lausanne,  Switzerland}%
\author{Marjaneh Jafari Fesharaki}
\affiliation{Department of Physics, Payame Noor University, PO Box 19395-3697 Tehran, Iran}%
\author{Henrik M. Ronnow}
\affiliation{Institute  of  Physics,  \'Ecole  Polytechnique  F\'ed\'erale  de  Lausanne  (EPFL),  CH-1015  Lausanne,  Switzerland}%
\author{Hossein Ahmadvand}
\affiliation{Department of Physics, Isfahan University of Technology, Isfahan 84156-83111, Iran}%
\date{\today}
\begin{abstract}
The magnetic, dielectric and structural properties of type-I multiferroic \knp~have been investigated by neutron powder diffraction, magnetization, dielectric and high temperature synchrotron-XRD measurements. Below the N\'{e}el transition of $T_\mathrm{N}$ = 25 K, KNiPO$_4$ displays a weakly non-collinear antiferromagnetic (AFM) structure with the orientation of the Ni$^{2+}$ magnetic moments mainly along $a$ axis. The compound crystallizes in the polar orthorhombic $Pna2_1$ space group at room temperature. A second-order structural phase transition corresponding to the onset ferroelectricity is observed at around $T_\mathrm{C}\sim$ 594(3)$^\circ$C, above which the crystal structure of \knp\, adopts the centrosymmetric $Pnma$ space group. The compound also displays another structural phase transition at $T_\mathrm{0}\sim$ 469 -- 488$^\circ$C, with a first-order character, which is attributed to the rearrangement of oxygen ligands, resulting in a change in the nickel ion co-ordination from four to five.
\end{abstract}
\maketitle
\section{Introduction}
Multiferroic materials, which display a strong coupling between magnetic and ferroelectric order, have received sustained scientific interest over the last decade and hold considerable promise for a wide range of technological applications. In these magnetoelectric multiferroics, time-reversal symmetry (TRS) and spatial inversion symmetry (IS) are spontaneously broken and a magnetoelectric coupling between the magnetic and ferroelectric orders is possible. Considering the prevalence of materials with either magnetic or ferroelectric order in devices (e.g. ferroelectric capacitors, transducers and actuators and ferromagnetic sensors and permanent magnets), using materials where both orders are coupled is the next natural step in technological development \cite{+2021}.

An example of a magnetoelectric multiferroic material where the magnetic and ferroelectric orders are coupled is \knp~\cite{Kurkin_Leskovets_Nikolaev_Turov_Turov_2003,Leskovets_Kurkin_Nikolaev_Turov_2002}. At room temperature, the orthorhombic unit cell of \knp\, can be described by the non-centrosymmetric $Pna2_1$ space group, where the broken IS gives rise to a spontaneous ferroelectric order~\cite{Fischer_Lujan_Kubel_Schmid}. On other hand, the Ni magnetic sub-lattice displays long-ranged magnetic order below $T_\mathrm{N}$ = 25\,K, signalling the breaking of TRS~\cite{Fischer_Lujan_Kubel_Schmid}. Lujan \textit{et al.} \cite{Lujan_Rivera_Kizhaev_Schmid_Triscone_Muller_Ye_Mettout_Bouzerar_1994} reported that below $T_\mathrm{N}$, certain off-diagonal magnetoelectric tensor elements of \knp\, are non-zero (namely, $\alpha_{21}$ and $\alpha_{12}$), indicating that the magnetic and ferroelectric orders in \knp\, are coupled. However, some questions regarding the nature of TRS and IS breaking remain, which we consider in turn.

In a powder neutron diffraction study of \knp\, at $T$\,=\,9\,K, Fisher \textit{et al.} proposed that the nickel ions display an antiferromagnetic order (with $\Gamma_1$ symmetry, see table \ref{tab:momentIrrep}) \cite{Fischer_Lujan_Kubel_Schmid}, where the components of the moments within each unit cell are fully compensated in all directions. This irreducible representation however, is not compatible with the magnetization measurements presented in Refs. \cite{Fischer_Lujan_Kubel_Schmid, Lujan_Rivera_Kizhaev_Schmid_Triscone_Muller_Ye_Mettout_Bouzerar_1994}, which suggest the presence of weak ferromagnetism in the compound. A theoretical study aimed at reconciling this inconsistency by discussing possible explanations for this phenomenon soon followed but no definite consensus was established~\cite{turov1996}.

High temperature transitions were found by Lujan \textit{et al.} \cite{High_Temp_KNIPO4,Lujan_Schmid_TissoT_1997, Lujan_Kubel_Schmid_1995} at $T_\mathrm{0}\sim$ 491$^\circ$C and $T_\mathrm{C}$ = 583(3)$^\circ$C using X-ray powder diffraction (XRD), polarized light microscopy and oscillating DSC techniques. The transition at  $T_\mathrm{C}$ was attributed to a symmetry lowering of the crystal structure of \knp\, from the $Pnma$ space group, which contains the inversion symmetry element, to the non-centrosymmetric $Pna2_1$ space group on cooling \cite{High_Temp_KNIPO4}. However, the crystal structure obtained from the refinement above $T_\mathrm{C}$ was not presented in the report.

As seen above, there has been a considerable effort to characterize \knp; however, there are still outstanding questions that merit further investigation. In this paper, we aim to characterize the magnetic structure of \knp\, at low temperatures. Specifically, high resolution neutron diffraction was used to determine the true ground state magnetic structure of the compound at $T$ = 1.5 K. Furthermore, synchrotron X-rays was used to determine the evolution of the size of unit cell and relative atomic positions of \knp, through the various high temperature transitions, namely $T_\mathrm{0}$ and $T_\mathrm{C}$, up to 750$^\circ$C, in order to confirm the spatial IS breaking proposed in Ref.~\cite{High_Temp_KNIPO4}.

Our high resolution single crystal diffraction results demonstrate that the crystal structure of \knp\, has the $Pna2_1$ space group at room temperature, consistent with previous reports. Magnetization measurements indicate a transition to an antiferromagnetic state below $T_\mathrm{N}$ and powder neutron diffraction reveals that the ground state magnetic structure of \knp\, is best described by the $\Gamma_1$ irreducible representation (irrep). The refinement of neutron data also shows that the effective moment of the Ni$^{2+}$ ions is mainly along the \textit{a} direction, with some canting along \textit{b} and \textit{c}. Nonetheless, for the $\Gamma_1$ irrep, the relative orientation of four nickel ions within each unit cell are fully compensated in all three directions. The refinement of our high temperature synchrotron results show that all but two atoms (namely, oxygen) in the unit cell adopt high symmetry positions consistent with the $Pnma$ space group above $T_\mathrm{C}$. We have also observed a large anomaly in the temperature dependence of the dielectric susceptibility of \knp\, at $T_\mathrm{0}$, associated with a change in Ni-O co-ordination, from five to four ($T>T_\mathrm{0}$). 

\section{Exprimental}
\label{sec:Experimenal}
The powders of KNiPO$_4$ was synthesized by a wet chemical method. Stoichiometric amounts of high purity Ni(NO$_3$)$_2$.6H$_2$O, C$_2$H$_3$KO$_2$ and H$_6$NO$_4$P were dissolved in deionized water and heated to 380 $^{\circ}$C with continuous stirring. After few hours, the mixture turns into a sticky yellow solid. The as-prepared powder was grounded and calcined at 400$^{\circ}$C for 24 hours, then grounded again and heated at 780$^{\circ}$C for 48 hours. All heated treatments were carried out in ambient atmosphere. 

To ascertain the quality and purity of our polycrystalline \knp\, sample, room temperature powder XRD measurements on a powder diffractometer (Empyrean, Malvern Panalytical) were performed with a Cu-$\kappa_{\alpha1}$ source ($\lambda\,=$\,1.5406\,\AA). To further refine the crystal structure, we isolated a single grain of \knp\, and performed single crystal XRD on a six-circle diffractometer (Rigaku) with a Mo-$\kappa_{\alpha}$ source ($\lambda\,=\,0.7107$\,\AA) at room temperature. Magnetization measurements were performed on a Quantum Design Physical Properties Measurement System (PPMS). Temperature-dependent magnetic susceptibility measurements were performed in the range of $T$ = 3 to 400 K, with zero-field cooled (ZFC) and field cooled (FC) protocols, in a fixed magnetic field of $H$=1 kOe. Field dependent magnetization measurements were performed up to 140 kOe at several fixed temperatures of $T$ = 3, 10, 20, 30\,K.

To determine the magnetic order below the N\'{e}el temperature, powder neutron diffraction was performed on polycrystalline \knp\, on the HRPT diffractometer at the Swiss Spallation Neutron Source. The incident neutron wavelength of 1.89\,\AA{} was selected with a Ge monochromator. High-statistics measurements were performed in the scattering angle ($2\theta$) range from 18$^\circ$ to 160$^\circ$, above and below $T_\mathrm{N}$, at $T$ = 30 and 1.5\,K with long counting times ($\sim$ 9 hours).

The dielectric constant of \knp\, was measured with a LCR meter (HIOKI-3536) and a programmable furnace in the temperature range of $T$ = 40 - 700$^{\circ}$C at a frequency of 1.5 MHz.

To uncover the evolution of the crystal structure in the vicinity of the high-temperature transitions, we performed synchrotron XRD on polycrystalline \knp\, on the MS-Powder beamline at the Swiss Light Source. The incident wavelength of $\lambda=$ 0.7085 \AA, was selected with a double crystal monochromator comprising two Si $(111)$ crystals. Measurements were performed over a temperature range from $T$ = 350 to 650$^{\circ}$C in 100$^{\circ}$C steps. The second series of high temperature synchrotron XRD measurements ($\lambda=$ 0.68306 \AA) were performed at the Swiss-Norwegian beamline (BM01) at the ESRF between $T$ = 30 and 750$^{\circ}$C, with 2$^{\circ}$C steps. The obtained data sets were refined with the Mag2Pol software\cite{Qureshi:in5014}. 

\section{Results and discussions}
\subsection{Structural characterization}
The analysis of XRD data, performed on both polycrystalline and an isolated single crystal grain of \knp, indicate that the compound has an orthorhombic crystal structure that can be described by the $Pna2_1$ space group at room temperature. The refinement of the positions of the atoms within the unit cell and the cell parameters, where $a$ = 8.6487(5)$\text{\AA}$, $b$ = 9.2778(5)$\text{\AA}$ and $c$ = 4.9154(5)$\text{\AA}$, are fully consistent with previous reports~\cite{Fischer_Lujan_Kubel_Schmid,Lujan_Kubel_Schmid_1995}.

\subsection{Magnetic structure}
\begin{figure}[b!]
    \centering
    \includegraphics[width=\linewidth]{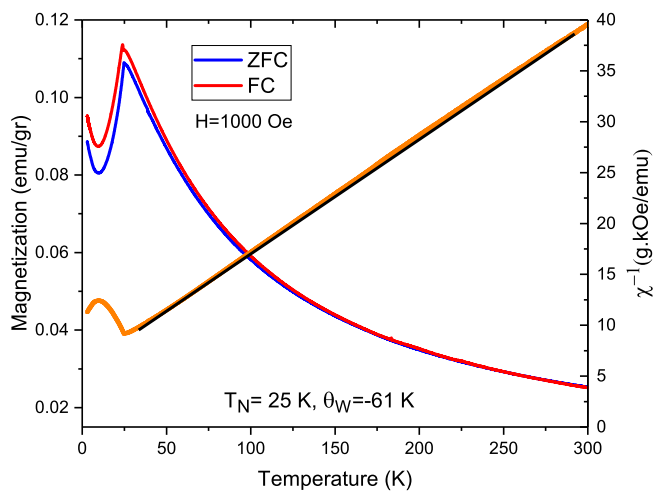}
    \caption{FC/ZFC magnetization (at 1000 Oe) and inverse magnetic susceptibility ($\chi^{-1}$) of \knp\, as a function of temperature. The ZFC data was fitted with a linear function in the temperature range above $T_\mathrm{N}$.}
    \label{fig:KNiPO4_MT}
\end{figure}

Figure~\ref{fig:KNiPO4_MT} plots the magnetization of \knp\, as a function of temperature. The FC and ZFC curves display an anomaly at $T_\mathrm{N}$ = 25(1)\,K, which indicates the onset of long-ranged antiferromagnetic order on the Ni magnetic sub-lattice.
To determine the extent to which the magnetic susceptibility above $T_\mathrm{N}$ follows the Curie-Weiss behavior, $C/(T-\theta_\mathrm{W})$, we also plotted the inverse magnetic susceptibility with respect to temperature in Fig.~\ref{fig:KNiPO4_MT}. As seen, the $\chi^{-1}$ curve is linearly proportional to $T$, with a Weiss temperature of $\theta_\mathrm{W}$= -61 K, indicative of strong antiferromagnetic interactions. Moreover, from the slope of the line, $C^{-1}$, we can obtain an estimate of the effective moment of Ni$^{2+}$ ions. Here, $C$ is defined as:
\begin{equation}
    C = \frac{n\mu_0 \mu_\mathrm{eff}^2}{3k_\mathrm{B}},
    \label{eq:mueff_C}
\end{equation}
where $n$ is the density of magnetic ions, $\mu_0$ is the vacuum permittivity and $k_\mathrm{B}$ is the Boltzmann constant. The effective moment of the Ni$^{2+}$ ions is estimated to be $\mu_\mathrm{eff}$ = 3.27 $\mu_\mathrm{B}$. The magnitudes of the nickel magnetic moment ($3d^8$) assuming un-quenched and fully quenched orbital contribution are given by $g_J[J(J+1)]^{1/2}=5.59$ and  $2[S(S+1)]^{1/2}=2.83$ $\mu_\mathrm{B}$ respectively, where $g_J$, is the Lande g-factor, $J$ and $S$ are the total and spin angular momentum. Hence, we deduce that the crystal environment in which the Ni ions reside, namely the $4a$ Wyckoff position in the $Pna2_1$ space group, does not fully quench the orbital angular momentum of $L=3$.

\begin{figure}[t!]
    \centering
    \includegraphics[width=1\linewidth]{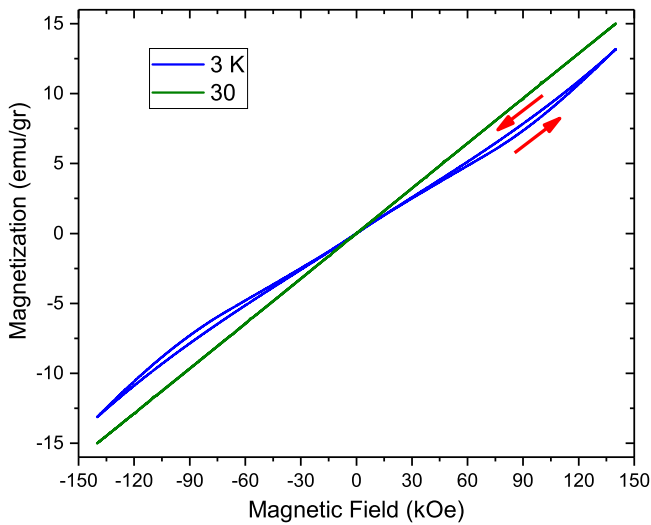}
    \caption{Magnetization of \knp~as a function of magnetic field (up to 140 kOe) at $T$ = 3\,K (below $T_\mathrm{N}$) and 30\,K (above $T_\mathrm{N}$). The arrows indicate the direction in which the measurements were performed.}
    \label{fig:KNiPO4_MH}
\end{figure}

Figure~\ref{fig:KNiPO4_MH} shows the magnetization of KNiPO$_4$ with respect to the applied magnetic field, up to 140 kOe, at two fixed temperatures of $T$ = 3\,K (below $T_{N}$) and 30\,K (above $T_{N}$). As expected for the paramagnetic phase, KNiPO$_4$ exhibits a linear $M$-$H$ dependence at $T$ = 30\,K. In the AFM phase, the $M$-$H$ curve deviates from linear behavior, around 70 kOe, with a small hysteresis. This behavior can be attributed to a spin-flop mechanism. The slight hysteresis indicates the first order nature of this mechanism. No saturation is seen up to 140 kOe, indicating that the magnetic structure remains in the canted spin flop state.   
\begin{figure}[t!]
    \centering
    \includegraphics[width=1.1\linewidth]{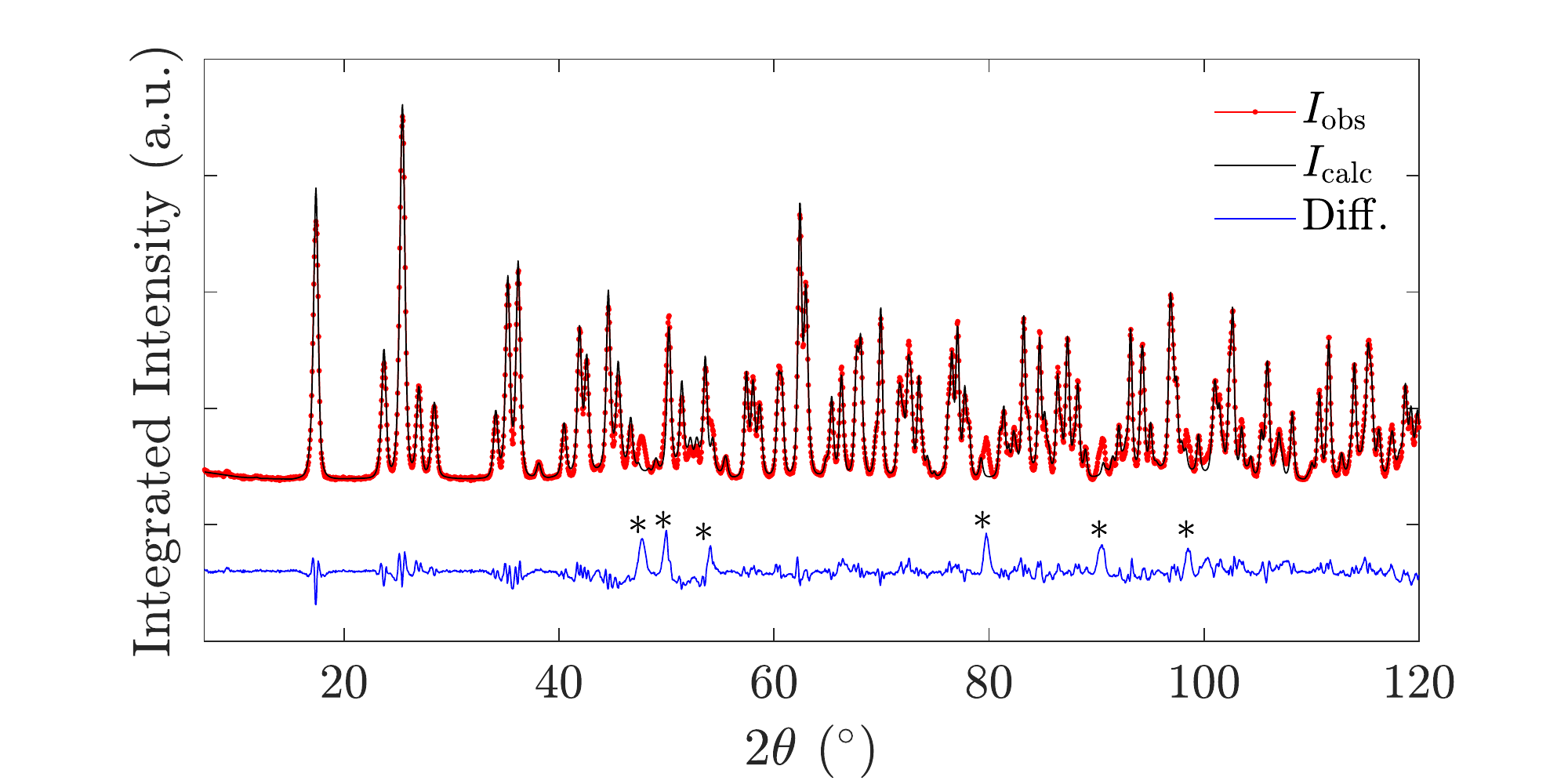}
    \caption{The neutron diffraction pattern of \knp\, (red) measured at $T$ = 1.5\,K. The blue curve indicates the difference between the measured and calculated ($I_\mathrm{calc}$) pattern obtained from the refinement. The asterisk $(\ast)$ denote scattering arising from an impurity phase.}
    \label{fig:KNiPO4_1K}
\end{figure}
\begin{table}[b]
	\caption{\label{tab:momentIrrep}
		Representational analysis and magnetic structure refinement for the $4a$ Wyckoff site of the $Pna2_1$ space group ($\textbf{k}=\textbf{0}$). The first column shows the irrep, the second column shows the spin orientation while the third column shows the relative spin arrangement at the atomic positions $(x,y,z)$, $(-x,-y,z+0.5)$, $(x+0.5,-y+0.5,z)$, and $(-x+0.5,y+0.5,z+0.5)$, respectively. The last two columns are tabulated the moments and the $\chi^2$ values of the refinement, respectively.}
	\begin{ruledtabular}
		\begin{tabular}{lcccr}
			Irrep&Orien-&Basis& $\mu_i$ & $\chi^2$\\
			&tation&vector  & ($i$ = $x$, $y$, $z$) & \\[2pt]
			\colrule
			&$x$&$(+--+)$& 3.93 \\
			$\Gamma_1$ &$y$&$(+-+-)$ & -0.36 & 53\\
			&$z$&$(++--)$ & 0.45\\
			\hline
			&$x$&$(+-+-)$ & 0.79\\
			$\Gamma_2$ &$y$&$(+--+)$ & -1.38& 58\\
			&$z$&$(++++)$&1.33\\
			\hline
			&$x$&$(++--)$&0.82\\
			$\Gamma_3$ &$y$&$(++++)$ & 0.74& 58\\
			&$z$&$(+--+)$&1.12 \\
			\hline
			&$x$&$(++++)$&1.62\\
			$\Gamma_4$ &$y$&$(++--)$ &1.79 & 58 \\
			&$z$&$(+-+-)$& 0.0\\
		\end{tabular}
	\end{ruledtabular}
\end{table}

\begin{figure}
	\centering
	\includegraphics[width=0.9\linewidth]{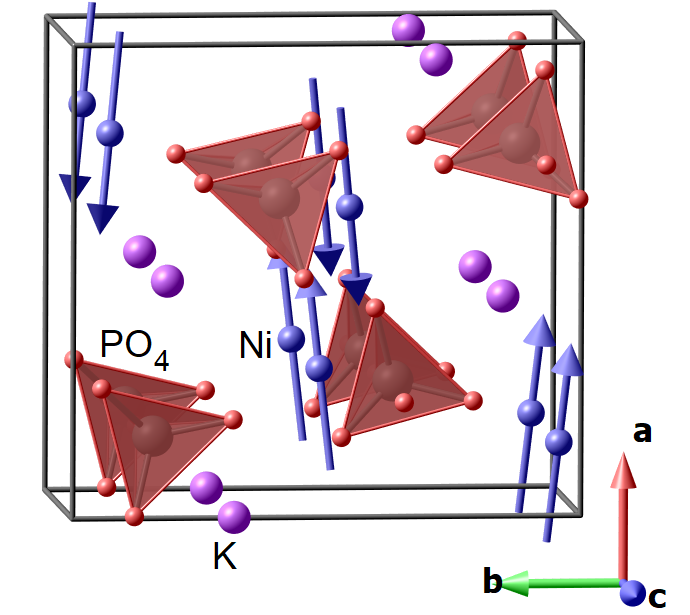}
	\caption{The non-collinear magnetic structure of \knp, obtained from the refinement of neutron diffraction pattern measured at $T$ = 1.5 K, can be described by the $\Gamma_1$ irrep with the main component of the Ni moments along the crystal $a$ axis.}
	\label{fig:KNIPO4_Moments}
\end{figure}

To determine the ground state magnetic structure of \knp, we now turn to consider the neutron diffraction measurement. Fig.~\ref{fig:KNiPO4_1K} shows the measured neutron diffraction pattern obtained at $T$ = 1.5 K, which is well below $T_\mathrm{N}$. The observed magnetic peaks can be indexed by the $Pna2_1$ space group, implying that the magnetic propagation vector is \textbf{k} = 0, where the magnetic structure is commensurate with the crystal structure of \knp. As such, the symmetry-allowed spin-configurations of the Ni magnetic sub-lattice, can be decomposed into 4 different irreducible representations (irreps), $\Gamma$ = $\Gamma_1+\Gamma_2+\Gamma_3+\Gamma_4$. All of the four irreps are three dimensional irreps, with three basis vectors each (see Table \ref{tab:momentIrrep}). For each of the 4 irreps, we refined the magnitude of the moments along all three directions.
Table~\ref{tab:momentIrrep} compares the goodness-of-fit parameters obtained from the refinement of the various irreps to the measured neutron data, from which we find that the $\Gamma_1$ irrep provides the best fit. This result corroborates the previously obtained results of Fischer \textit{et al.}~\cite{Fischer_Lujan_Kubel_Schmid} which found the spin configuration of the nickel ions of \knp~to follow the same representation. The refined moment is $\mu = 3.87 \pm 0.14 [\mu_B]$ which is consistent with the magnetization measurements. Indeed we find that for the $\Gamma_1$ irrep, all of the basis vector components along ($x$, $y$, $z$) are compensated, which is consistent with the temperature dependent magnetization measurement (Fig. \ref{fig:KNiPO4_MT}). The magnetic structure of \knp, illustrated in Fig.~\ref{fig:KNIPO4_Moments},  display $\Gamma_1$ symmetry with the Ni magnetic moments predominantly oriented along $a$ axis and exhibit a weakly non-collinear AFM structure.

\subsection{Synchrotron-XRD Experiment}
\begin{figure}[t!]
    \centering
    \includegraphics[width=1.0\linewidth]{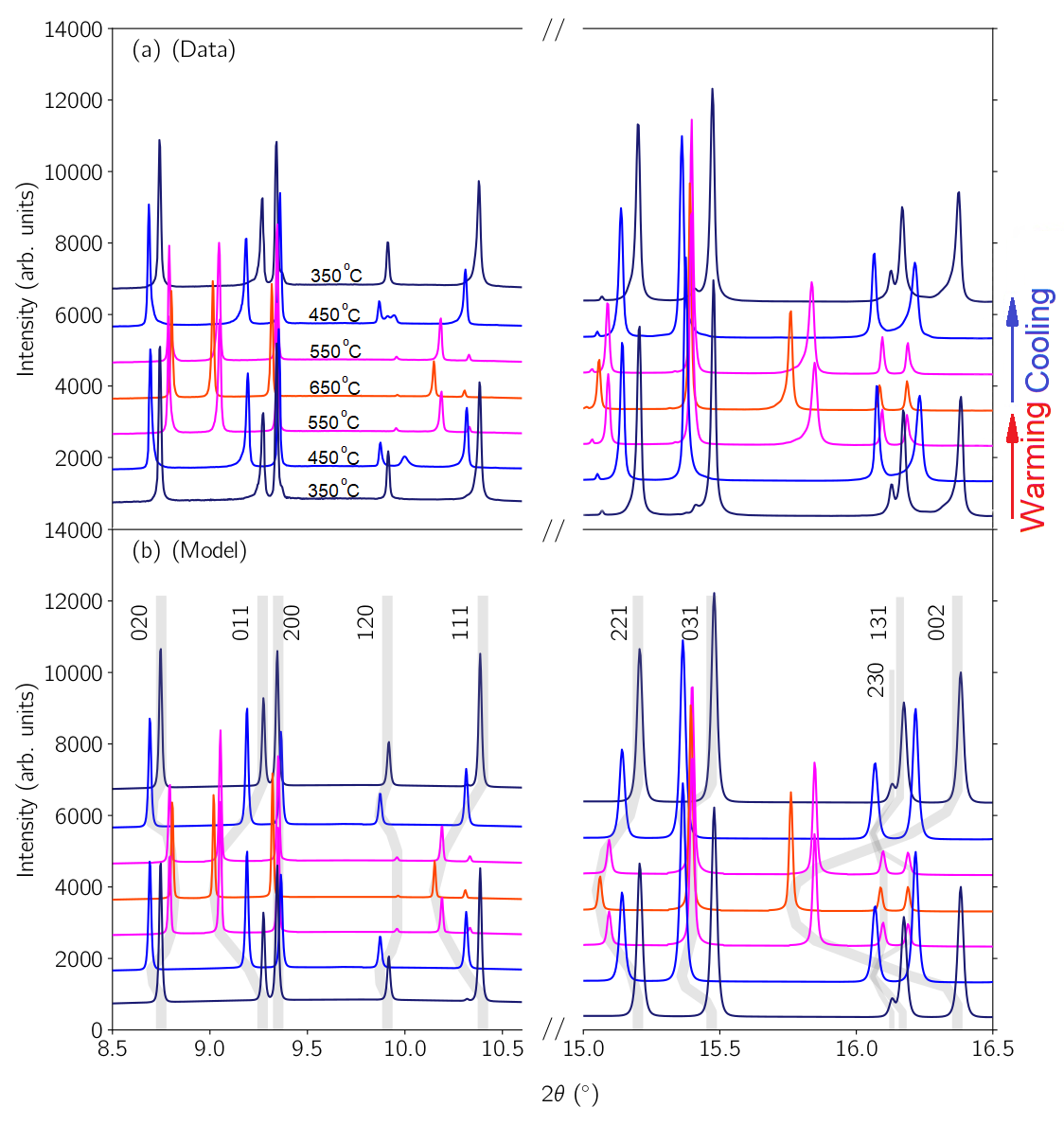}
	\caption{Comparison between the (a) measured and (b) calculated XRD patterns of \knp\, at various temperatures. The temperature evolution of the crystal axes, $a$, $b$ and $c$, are denoted by the (200), (020), and (002) reflections, respectively. The changes in the relative intensities of the peaks account for the reorganization of the atomic positions within the unit cell. The gray curves follow the indexed peaks over the temperature range.}
	\label{fig:Xray_Data_Zoom}
\end{figure}

While the Li-based orthophosphates, Li$M$PO$_4$ ($M$ = Co, Fe, Ni) belong to family of type-II multiferroics \cite{Gnewuch,Petersen}, \knp~can be considered as a type-I multiferroic, in which the ferroelectric and magnetic orders originate from different sources. Thus, in order to clarify structural-related phenomena, we now turn to consider high-temperature synchrotron XRD measurements. In particular, we want to (i) study how the crystal structure and cell parameters change with temperature and (ii) uncover the evolution of the atomic positions within the unit cell. The XRD patterns were collected on the MS-Powder beamline from $T$ = 350$^\circ$C to 650$^\circ$C and back to 350$^\circ$C, in 100$^\circ$C steps. To draw out salient features of the temperature dependence, we plotted the measured data in Fig.~\ref{fig:Xray_Data_Zoom}(a) for the $2\theta$ ranges: 8.5$^\circ$--10.5$^\circ$ and 15.0$^\circ$--16.5$^\circ$. The arrows indicate the warming and cooling sequences. 

\begin{figure*}[t!]
\centering
\begin{minipage}{0.7\textwidth}
  \centering
  \includegraphics[width=1.0\linewidth]{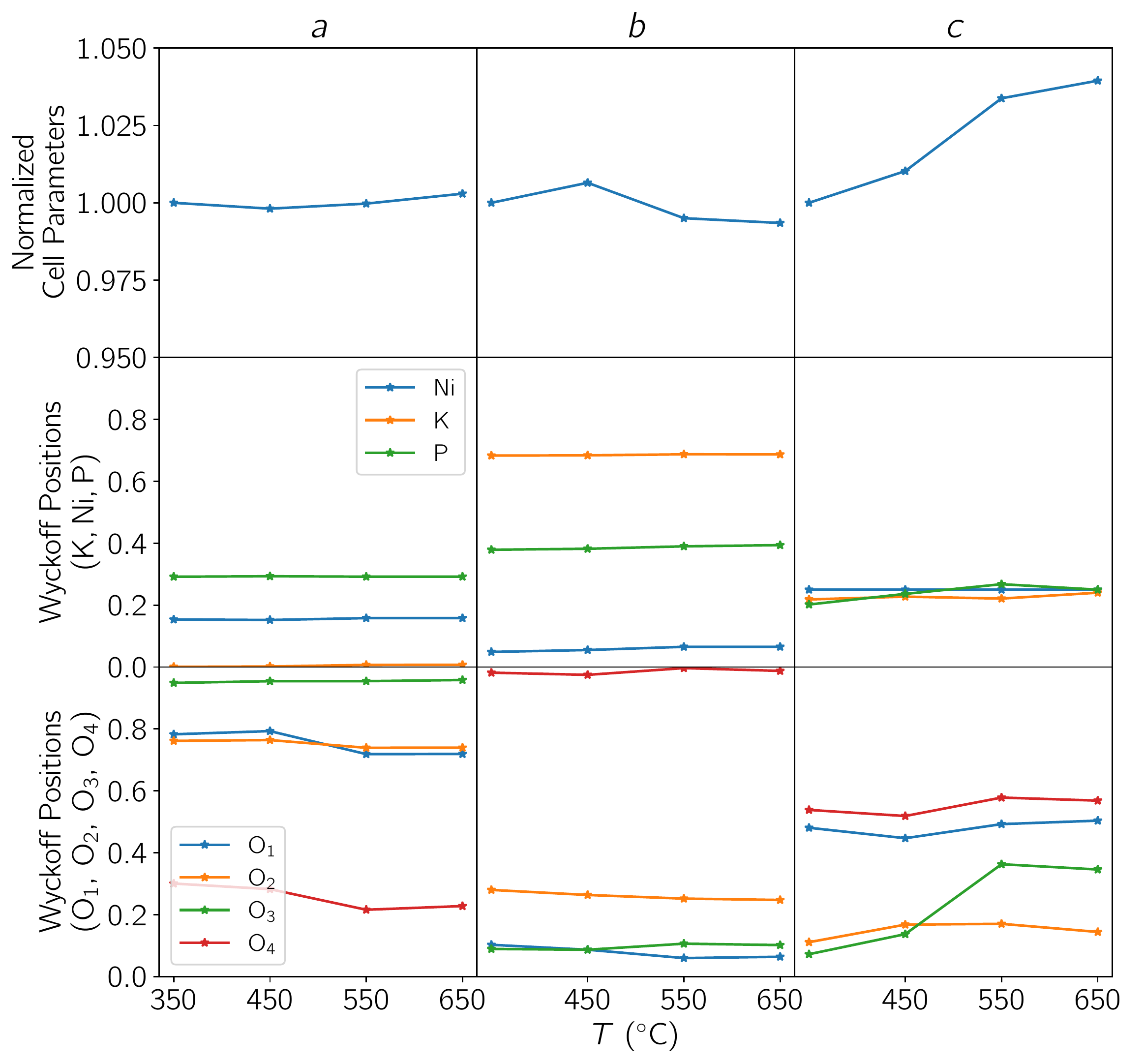}
\end{minipage}%
\begin{minipage}{0.3\textwidth}
  \centering
  \includegraphics[width=0.8\linewidth]{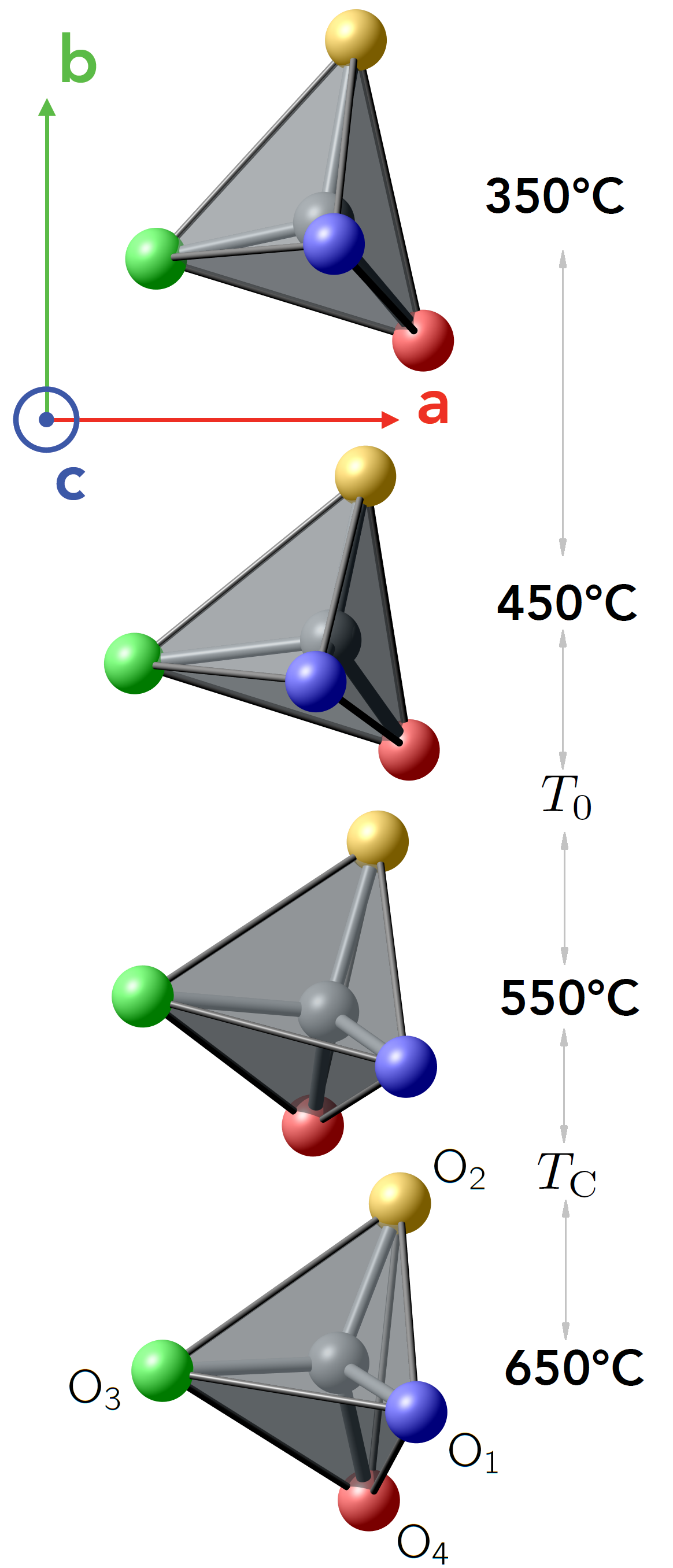}
\end{minipage}
\caption{(Left) Top row: temperature evolution of the cell parameters of \knp~normalized relative to the value at $T$ = 350$^\circ$C. Middle row: the $x$,$y$, and $z$ Wyckoff positions of K, Ni and P atoms. Bottom row: Wyckoff positions of the four symmetry in-equivalent oxygen atoms. (Right) Evolution of the PO$_4$ tetrahedron with temperature with four in-equivalent oxygen positions labelled O$_1$, O$_2$, O$_3$ and O$_4$.}
\label{fig:Atom_pos}
\end{figure*}

Here, we only consider the refinement of the data on warming. The $Pna2_1$ space group was used in the refinement of the data. The calculated XRD patterns are illustrated in Fig.~\ref{fig:Xray_Data_Zoom}(b), with the gray curves highlighting the evolution of the indexed peaks with temperature. Significant shifts in peak position are observed, especially the peaks (020) and (002) which changes drastically compared to the (200) reflection. Hence, this refinement result allows us to infer on the evolution of the cell parameters: while \textit{a} remains relatively constant, \textit{b} decreases slightly and \textit{c} shows the most drastic changes as it increases with temperature [See Fig.~\ref{fig:Atom_pos}, top row]. Besides shifts along 2$\theta$, our model also describes the changes in intensity of the peaks, especially of the (111) and (221) reflections [See Fig.~\ref{fig:Xray_Data_Zoom}(b)], indicating movement of the atoms within the unit cell.

As previously mentioned, refinements of the synchrotron data were performed with the choice of the $Pna2_1$ space group. However, Lujan \textit{et al.} \cite{High_Temp_KNIPO4} proposed that the high temperature structure of KNiPO$_4$ could be isostructural to CsZnPO$_4$, which crystallizes in the $Pnma$ space group. Many similarities can be drawn between the $Pna2_1$ and $Pnma$ space groups: the $Pnma$ space group has eight symmetry elements, generated by the identity, a $2_1$ screw axis, a glide transformation and an inversion; the $Pna2_1$ space group, on the other hand, lacks the inversion and therefore only has four symmetry elements. For \knp\, to gain an inversion center, all of the atoms need to simultaneously adopt high symmetry positions that satisfy both space groups, such as any of the following: 
\begin{multline}
\label{eq:highsympos}
    (x,y,z) = (0,0,0), (0,1/2,0), (x,y,1/4),  (\pm x,\pm y,\pm z).
\end{multline}
To test the hypothesis that \knp\, gains an inversion center, we propose a two step workflow.
The first step is to determine the relative rearrangement of K, Ni, P and the four symmetry inequivalent oxygen atoms within the unit cell. The second step is to identify a suitable global spatial translation ($x_0$, $y_0$, $z_0$) for all of the atoms in the $Pna2_1$ space group. 
This is owing to the fact that the $Pna2_1$ space group does not possess a unique origin, since:
\begin{equation}
    f(x_k, y_k, z_k) = f(x_k+x_0,y_k+ y_0,z_k+ z_0).
\end{equation}
for any global spatial translation, 
\begin{equation}
    (x_0, y_0, z_0) \in \mathbb{R}^3,
\end{equation}
where $f$ refers to any function (Hamiltonian, eigenfunctions, etc.) that transforms with the symmetry elements of the $Pna2_1$ space group.

If the atoms within the unit cell do indeed rearrange such that they adopt the high symmetry positions in Eq. \ref{eq:highsympos} given a suitable choice of the global offset ($x_0$,$y_0$,$z_0$), then the compound gains the inversion symmetry element \footnote{However, due to the different order of symmetry operations between $Pna2_1$ and $Pnma$, the \textit{y} and \textit{z} positions are swapped.}. This is a very attractive proposition to account for the ferroelectricity driven by the IS breaking. 

We adopted this two-step workflow and plotted the positions of the atoms with respect to temperature (Fig. \,\ref{fig:Atom_pos}). As temperature increases from $T$ = 350 to 650$^\circ$C, the potassium, nickel and phosphorus atoms organize themselves in two planes (at $z$ = 1/4 and $z$ = 3/4) which can be mapped onto the 4$c$ Wyckoff site of the $Pnma$ space group \footnote{Noting that the $y$ and $z$ coordinates are swapped.}. Indeed, by symmetry the $x$ and $y$ coordinates can take any value and it is mainly the $z$ Wyckoff position that is crucial (see Eq.~\ref{eq:highsympos}). This might account for the fact that the $c$ parameter of the unit cell of \knp\, changes drastically [See Fig.~\ref{fig:Atom_pos}, top row], while $a$ and $b$ remain relatively constant. Similarly, the $O_1$ and $O_4$ pair seems to tend towards the high-symmetry points $(\pm x,\pm y,\pm z)$, where $O_1$ and $O_4$ map onto each-other via the inversion symmetry element.

Here we demonstrate that the K, Ni, P, O$_1$ and O$_4$ atoms can adopt the high-symmetry positions of the $Pnma$ space group. However, the $O_2$ and $O_3$ pair seems to organize themselves in the same high-symmetry plane [see Fig.~\ref{fig:Atom_pos}, bottom row], albeit with large deviations.

\begin{figure}
		\centering
	\includegraphics[width=1\linewidth]{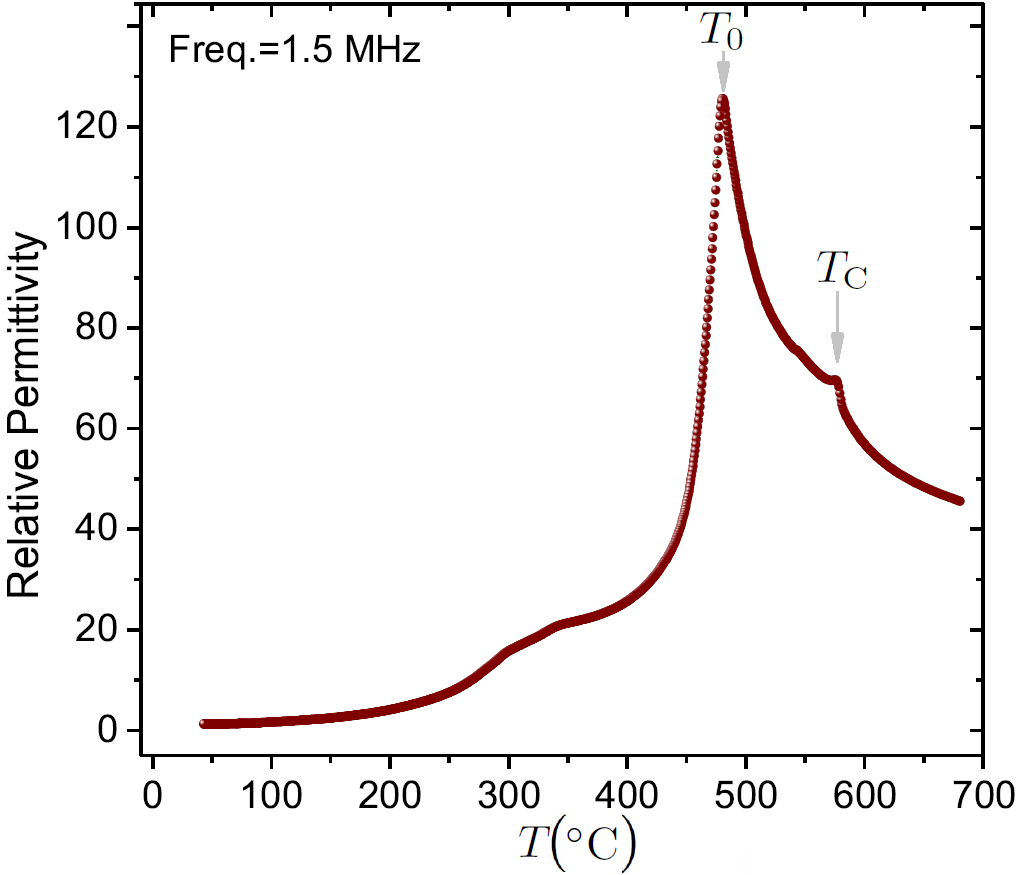}
	\caption{Temperature dependence of relative permittivity of \knp~at a frequency of 1.5 MHz.}
	\label{fig:perm}
\end{figure}
\begin{figure*}
	\centering
	\includegraphics[width=1.0\linewidth]{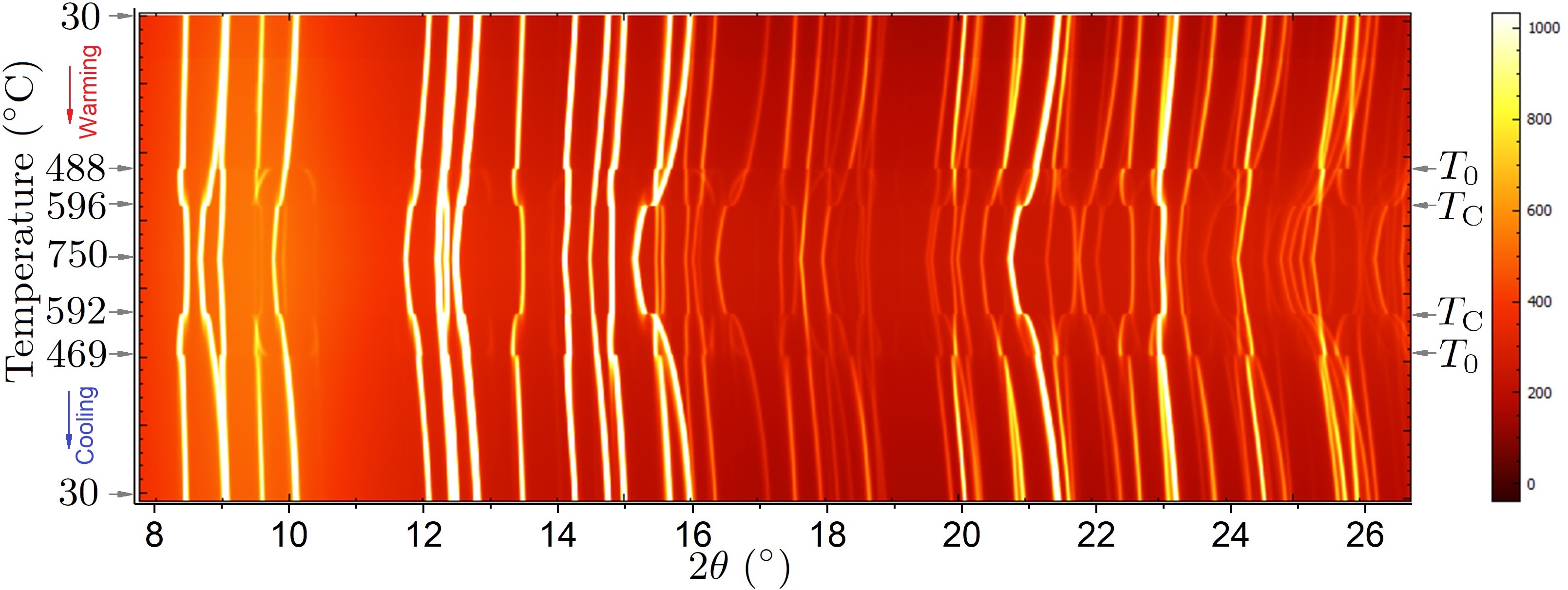}
	\begin{minipage}{1.0\textwidth}
	\centering
	\includegraphics[width=0.9\linewidth]{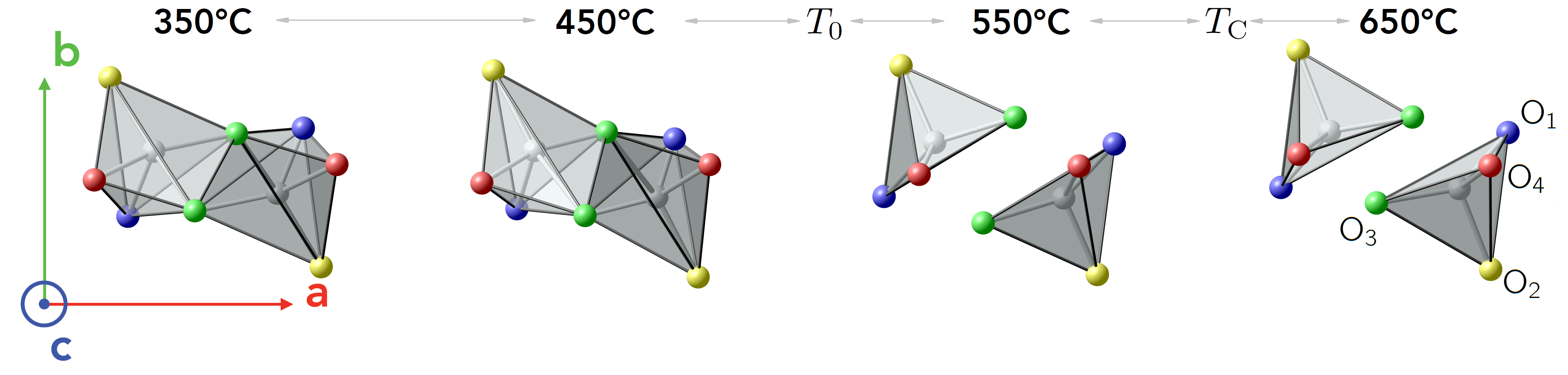}
\end{minipage}
	\caption{ (Top) Contour plot of the synchrotron-XRD patterns of \knp, measured during warming from 30$^\circ$C  to 750$^\circ$C  and again during cooling to 30$^\circ$C, with 2$^\circ$C step. The transition temperatures $T_\mathrm{0}$ and $T_\mathrm{C}$ are denoted on the right axis. (Bottom) Evolution of Ni polyhedron with temperature.}
	\label{fig:scan}
\end{figure*}
We also find that the size of the PO$_4$ tetrahedron (namely the P - O bond length) is fairly constant. PO$_4$ rotate about \textit{a} and \textit{b} but marginally about \textit{c} [See Fig.~\ref{fig:Atom_pos} (Right)]. Two atoms adopt position close to $8d$ consistent with the hypothesis, while two others draw relatively close to the predicted positions but not enough to fully adopt the high-symmetry site.

As discussed above, various structural changes occur in \knp. Dielectric permittivity measurements are complementary tool to investigate such temperature-induced structural changes. Fig.~\ref{fig:perm} shows the relative permittivity of \knp\, as a function of temperature from $T$ = 40 -- 680$^\circ$C. The value of the dielectric permittivity is small, relative to other ferroelectrics. Most importantly, two anomalies are seen at around $T_\mathrm{0}\sim$ 480$^\circ$C and $T_\mathrm{C}\sim$ 580$^\circ$C. These anomalies are in good agreement with the DSC measurements reported in earlier work \cite{Fischer_Lujan_Kubel_Schmid} on \knp~single crystals. To establish a precise connection between evolution of crystal structure and dielectric anomalies, we measured the XRD of \knp\, with small temperature interval of 2$^\circ$C. The measurements were performed from $T$ = 30--750$^\circ$C at Swiss-Norwegian beamline (BM-01). Fig.~\ref{fig:scan} shows a 2D contour plot of the collected patterns, from $T$ = 30--750$^\circ$C on warming and again to 30$^\circ$C on cooling. Anomalies in the XRD associated with the aforementioned structural transitions at $T_\mathrm{C}$ and $T_\mathrm{0}$ are clearly seen [See Fig.~\ref{fig:scan}]. The anomaly in the vicinity of $T_\mathrm{0}$ displays thermal hysteresis since it is observed at 488$^\circ$C and 469$^\circ$C during warming and cooling, respectively. On the other hand, the structural transition at $T_\mathrm{C}\sim$= 594(3)$^\circ$C displays close to no thermal hysteresis.

While nickel ions can, generally, have coordination numbers of 4, 5, and 6, the realization of five-coordination Ni$^{2+}$ are rare~\cite{Galoisy_Calas_1993}. As depicted in Fig.~\ref{fig:scan} (bottom), the Ni$^{2+}$ ion in \knp\, resides in a five-coordinated distorted square pyramid polyhedron at 350$^\circ$C. On warming, the atomic positions of the oxygen ligands are rearranged within the unit cell, which leads to the appearance of a clear transition at 488 $^\circ$C in the contour plot of XRD patterns, concomitant with the anomaly observed in the dielectric constant at $T_\mathrm{0}$ (Fig.\ref{fig:perm}). This anomaly is attributed to the fact that five-coordinated Ni$^{2+}$ in square pyramidal field changes to a tetrahedral environment [See Ni polyhedron at 650$^\circ$C in Fig. \ref{fig:scan} (bottom)] above 488$^\circ$C during warming, and vice versa below 469$^\circ$C during cooling. The thermal hysteresis indicates the first-order nature of this unique structural transition. 

A second structural transition observed in the contour plot at 594(3)$^\circ$C correspond to the anomaly at $T_\mathrm{C}$ in the temperature dependence of the permittivity of \knp. As deduced above from symmetry analysis, the high temperature phase above $T_\mathrm{C}$ belongs to the non-polar $Pnma$ space group, indicating that the transition at  $T_\mathrm{C}$ ($Pnma\leftrightarrow Pna2_1$) can be assigned to the onset ferroelectricity for \knp.

\section{Conclusion}
In this work, we present a comprehensive study of the magnetic and crystal structure of \knp. Magnetization measurements show an antiferromagnetic order below $T_\mathrm{N}$=25(1) K, with a slight spin-flop behavior at low temperatures. Neutron diffraction at $T$ = 1.5\,K, reveal a weakly non-collinear structure in the $Pna2_1$ space group, with magnetic propagation vector \textbf{k} = 0. The value of effective moment of Ni$^{2+}$ ions was found to be 3.27 and 3.87\,$\mu_\mathrm{B}$ from the magnetic susceptibility and neutron diffraction measurements, respectively, which state that the orbital momentum lies in between the unquenched and quenched states.

The crystal structure of \knp\, was studied by synchrotron radiation, in the temperature range of $T$ = 20 to 750$^{\circ}$C, in order to characterize the structural transitions. Refinement of the data suggests anisotropic changes in the unit cell of \knp, where the most drastic change was observed in \textit{c} lattice parameter. Although the $Pna2_1$ space group produces good fits to the measured diffraction patterns at all temperatures, using symmetry analysis we deduce that at high temperatures, the atomic positions seem to tend towards the centrosymmetric $Pnma$ space group. Using thermo-XRD and temperature dependence of permittivity two structural transitions were observed: the anomaly at $T_\mathrm{0}\sim$ 469--488$^{\circ}$C is associated with the rearrangement oxygen ligands of the Ni$^{2+}$ ions, going from a five-coordinated polyhedral to tetrahedral environment; the anomaly at around $T_\mathrm{C}\sim$ 590$^{\circ}$C is related to the structural phase transition of \knp\, with the crystal structure going from the $Pnma$ to  $Pna2_1$ space group which is associated with the onset of ferroelectricity.

\begin{acknowledgements}
This work was supported by ZHAW university, based on the bridging grant with South Asia and Iran. The beamline proposal numbers for the experiments are 20210820 (HRPT, SINQ, PSI and MS-POWDER, SLS, PSI) and 01-02-1247 (BM01, ESRF). We thank V. Pomjakushin, A. Cervellino, D. Chernyshov, C. McMonagle and M. Akhyani for assistance during the various experiments. JRS acknowledges support from the Singapore National Science Scholarship from the Agency for Science Technology and Research and the European Research Council under the European Union’s Horizon 2020 research and innovation program synergy grant (HERO, Grant No. 810451).
\end{acknowledgements}
\section*{References}
\bibliographystyle{apsrev4-2}
\bibliography{ref}
\end{document}